\documentclass[aps,prl,floats,twocolumn,twoside,floatfix]{revtex4-1}

\usepackage{graphicx}
\usepackage{amsmath,amssymb}
\usepackage{psfrag}
\usepackage{color}

\begin{document}

\title{Percolation approach to glassy dynamics with continuously
  broken ergodicity}

\author{Jeferson J. Arenzon$^1$}

\author{Antonio Coniglio$^2$}

\author{Annalisa Fierro$^{2,3}$}

\author{Mauro Sellitto$^4$}

\affiliation{$^1$Instituto de F\'{\i}sica, Universidade Federal do Rio Grande
  do Sul, CP 15051, 91501-970 Porto Alegre RS, Brazil}

\affiliation{$^2$Dipartimento di Scienze Fisiche, Universit\`a di Napoli
  ``Federico II'', \\ Complesso Universitario di Monte Sant'Angelo, Via
  Cintia, 80126 Napoli, Italy}

\affiliation{$^3$CNR-SPIN, Via Cintia, 80126 Napoli, Italy}

\affiliation{$^4$Dipartimento di Ingegneria Industriale e
  dell'Informazione, \\ Seconda Universit\`a di Napoli, Real Casa
  dell'Annunziata, I-81031 Aversa (CE), Italy }

\newcommand{\kB}{k_{\scriptscriptstyle \rm B}}
\newcommand{\Tg}{T_{\scriptstyle \rm g}}
\newcommand{\taue}{\tau_{\epsilon}}
\newcommand{\tc}{t_{\scriptscriptstyle \rm c}}
\newcommand{\ts}{t_{\scriptscriptstyle s}}
\newcommand{\dU}{d_{\scriptscriptstyle \rm U}}
\newcommand{\df}{d_{\scriptscriptstyle \rm f}}
\newcommand{\e}{\epsilon} \newcommand{\ph}{\varphi}
\newcommand{\pc}{p_{\scriptscriptstyle \rm c}}
\newcommand{\rhoc}{\rho_{\scriptscriptstyle \rm c}}
\newcommand{\phiEA}{\phi_{\scriptscriptstyle \rm EA}}
\newcommand{\M}{\mathsf M} \newcommand{\MCT}{{\small MCT}}

\date{\today}

\begin{abstract} 
We show that the relaxation dynamics near a glass transition with
continuous ergodicity breaking can be endowed with a geometric
interpretation based on percolation theory. At mean-field level this
approach is consistent with the mode-coupling theory (\MCT) of type-A
liquid-glass transitions and allows to disentangle the universal and
nonuniversal contributions to \MCT\ relaxation exponents. Scaling
predictions for the time correlation function are successfully tested
in the $F_{12}$ schematic model and facilitated spin systems on a
Bethe lattice.  Our approach immediately suggests the extension of
\MCT\ scaling laws to finite spatial dimensions and yields new
predictions for dynamic relaxation exponents below an upper critical
dimension of 6.
\end{abstract}

\pacs{} 

\date{\today}

\maketitle

Percolation~\cite{stauffer} is one of the most appealing examples of
phase transitions which has been successfully applied to a variety of
problems and has provided a deeper insight into the theory of critical
phenomena.  Although its relevance to amorphous magnets and structural
glasses has been often suggested, the geometrical interpretation of
scaling laws observed during glassy relaxation has been thwarted by
several difficulties and remains one of the fundamental issues of
condensed matter science.

In this paper, we formulate a percolation approach to glassy dynamics
with continuously broken ergodicity and exploit its predictions to
provide a novel interpretation of the mode-coupling theory (\MCT) of
continuous (or, type A) liquid-glass transition~\cite{Gotze} and to
suggest new scaling relations.  Two key questions lie at the heart of
our work: is there any universality --in the sense of critical
phenomena-- hidden in \MCT\ exponents? Is there any geometrical
picture underlying \MCT\ scaling laws?

We develop our formalism for a generic spatial dimension and in the
mean-field limit, but our approach can in principle be applied to a
range of different systems whenever clusters play an essential role.
The mean field limit will make transparent and explicit the connection
with \MCT\ and will naturally yield a finite-dimensional extension of
\MCT\ scaling laws.  Specifically, we generalize to finite dimensions
the universal scaling law relating the exponents of structural
relaxation time and critical decay law, and uncover a new intermediate
relaxation regime close to criticality. In this regime, the time
correlation function is a combination of algebraic and stretched
exponential decay, with precise predictions in terms of percolation
critical exponents, and new universal scaling relations.  Comparisons
with the analytical and numerical solution of the schematic
\MCT\ equation, and with Monte Carlo simulation of facilitated spin
models on a Bethe lattice give excellent results.

\paragraph{Schematic mode-coupling theory.---}

We first recall the \MCT\ result for the type A liquid-glass
transition. We focus on the schematic ${\mathsf F}_{12}$ model which
is known to reproduce the basic features of glassy
dynamics~\cite{Gotze}. In this case, the memory kernel functional $\M$
is:
\begin{eqnarray}
  \M [\phi(t)]  & = & v_1 \phi(t) + v_2 \phi^2(t),
  \label{F12}
\end{eqnarray} 
where $v_1,\, v_2$ are parameters controlling the system state and the
correlator $\phi(t)$ of density fluctuations at time $t$ obeys the
integro-differential equation:
\begin{eqnarray}
  \phi(t) + t_0 \, \dot \phi(t) + \int_0^t \M[\phi(t-s)] \ \dot{\phi}(s)\,
  \mathrm{d}s &=& 0,\label{MCT}
\end{eqnarray}
where $t_0$ is a characteristic microscopic timescale and overdamped
local motion is assumed for simplicity.  We are interested here in the
region of phase diagram in which the ergodic-nonergodic transition is
continuous.  Continuous glass transitions have recently attracted
significant attention in connection with the behavior of fluids
confined in porous media~\cite{Krak,KurzKim,Franosch}, which is related
to the diffusion-localization behavior of the Lorentz gas. In this
context the relevance of percolation was first noticed~\cite{GoLeYi}
(and~\cite{Go} for the related problem of the conductor-insulator
transition of quantum fluids in a random potential).

In the ${\mathsf F}_{12}$ schematic model the continuous transition
line corresponds to the segment $v_1=1$ and $v_2 \in [0,1]$. At
criticality relaxation is algebraic, $\phi(t) \sim t^{-a}$, and the
structural relaxation time $\taue$ at distance $\e$ from the critical
line behaves as $\taue\sim |\e|^{-\zeta}$ near the transition. The
exponents $a$ and $\zeta$ are not independent but connected throughout
the continuous glass transition range by~\cite{Gotze}:
\begin{equation} 
 \lambda = \frac{1}{2} \, \frac{\M''(q)}{(\M'(q))^{3/2}} , \qquad
 \zeta = \frac{1}{a}, \qquad \frac{\Gamma^2(1-a)}{\Gamma(1-2a)} =
 \lambda ,
  \label{eq.gamma}
\end{equation} 
where $\Gamma$ is the Euler's gamma function, the kernel derivatives
relative to $\phi$ are computed at criticality, and $\lambda$ is the
so-called parameter exponent. For the ${\mathsf F}_{12}$ schematic
model $\lambda=v_2/v_1^{3/2}$ and, in particular, $\lambda=v_2$ in the
continuous glass transition range.  In the frozen phase the
Edwards-Anderson order parameter, $q=\lim_{t \to \infty} \phi(t)$,
approaches zero as $q \sim \epsilon^{\beta}$ with $\beta=1$, which
coincides with the mean field value of the order parameter percolation
exponent. Following this hint we develop a dynamical percolation model
describing relaxation dynamics near the type A transition.

Before we proceed the rather peculiar nature of \MCT\ should be
emphasized: on one hand, it gives rise to relaxation patterns with
universal scaling relation between the exponents $\zeta$ and $a$; on
the other hand, these exponents must be considered nonuniversal,
because they depend on the kernel parameters.  Further, the
\MCT\ glass transition has a purely dynamical nature unrelated to any
equilibrium singularity.  These unconventional features are a major
stumbling block in understanding the glassy relaxation as an ordinary
dynamical critical phenomenon (see, however,
refs.~\cite{BiBoMiRe,sarl,SzFl,Franz,Parisi} for recent progress in
this direction).  In our cluster interpretation of \MCT\ we connect
the parameter $\lambda$ to the cluster lifetime and disentangle the
universal and nonuniversal features encoded in the exponents $a,\,
\zeta$ and $\lambda$. These features turn out to be related to
the underlying geometric structure of percolation clusters and to the
cluster lifetime. From this connection new universal scaling relations
valid in generic spatial dimensions will follow.

\paragraph{Percolation approach.---}

We first note that the \MCT\ relation $\zeta = 1/a$ can be generally
understood in terms of a simple scaling argument. Indeed, if we write
\begin{equation}
 \phi(t,\e) = \e^{\beta}  {\cal F} \left(\frac{t}{\taue}\right),
  \label{phi1}
\end{equation}
where the scaling function ${\cal F}(u)$ has limiting behaviors:
\begin{equation}
{\cal F}(u) \propto \left\{
\begin{array}{lcr}
u^{-\beta/\zeta} \,, & \mbox{for}  & u \ll 1 \\
{\rm cst.}  \,, &\mbox{for} & u \gg  1
\end{array}  
\right.
\label{phi3}
\end{equation}
then relation $\zeta=1/a$ is correctly recovered with $\beta=1$.

To explicitly obtain the scaling function we introduce a percolation
approach elaborating on the cluster formulation used to describe the
sol-gel transition~\cite{FiAbCo}.  We posit that our reference glassy
system can be described as a collection of clusters each of which
decays exponentially in time over a timescale $\ts$ that increases
with the cluster size $s$.  Structure and relaxation of clusters
evidently depend on the precise nature of the system interaction and
the underlying microscopic dynamics.  We shall assume a power-law
behavior of cluster lifetime, $\ts \propto s^{x}$, as born out from
experimental results and simulations on
polymers~\cite{theoryandexperimets}.
Under these hypothesis the correlator, $\phi(t)$, describing global
relaxation, can be written as the superposition of the decay of
different clusters:
\begin{equation}
  \phi(t) = \sum_s s \left. n(s) \, \exp\left(-\frac{t}{\ts} \right)
  \right/ \sum_s s \, n(s)
\end{equation}
where $n(s)$ is the cluster size distribution.  This expression can be
evaluated exactly on a Bethe lattice, by using the results of
Ref.~\cite{EssamFisher}. Alternatively, using the asymptotic
expression for the cluster size distribution~\cite{stauffer} one can
perform a saddle-point integration for a general $d$-dimensional
system near the percolation threshold~\cite{FiAbCo}.  Using the latter
method one finds the following relaxation regimes. In the fluid phase,
at distance $\e$ from the threshold, one has that the system relaxation is
described by 
a combination of algebraic and stretched exponential
decay~\cite{cola}:
\begin{equation}
  \phi(t,\e) \sim \e ^{\beta} \left(\frac{\taue}{t}\right)^{c}
  \exp\left[-\left(\frac{t}{\taue}\right)^y \right],
\label{phi_scaling}
\end{equation}
where exponents $c$ and $y$ are related to $x,\, \beta$ and $\gamma$
by~\cite{notax}:
\begin{equation}
 c = \frac{3 \beta +\gamma}{2(\beta + \gamma)(x+1)}, \qquad y =
 \frac{1}{x+1},
\label{c}
\end{equation}
The relaxation time and the critical decay law turn out $ \taue \sim
\left| \e \right|^{-\zeta}$ and $\phi(t)\equiv \phi(t,\e=0) \sim
t^{-a}$ where
\begin{equation}
\zeta= x \, (\beta + \gamma) , \qquad a=\frac{\beta}{x (\beta + \gamma)}.
\label{a}
\end{equation}
In the frozen phase, where ergodicity is broken due to the appearance
of a percolating cluster, the aymptotic value of the correlator is
$q\sim |\e|^{\beta}$, and the non-arrested part, $\phi(t,\e) - q$, has
the same form as Eq.~(\ref{phi_scaling}) with exponents $c'$ and $y'$
related to those in the fluid phase by
\begin{equation}
c'= \frac{c/y - 1/2d}{1/y-1/d}, \qquad y'= \frac{1-1/d}{1/y-1/d}.
\label{c_prime}
\end{equation}
Note that the above exponents depend on the microscopic mechanism
responsible for the single cluster relaxation, i.e., on $x$.
Nevertheless, one can suitably combine them to get universal scaling
laws that only depend on the percolation exponents. We shall come back
to this important point later in our comparison with \MCT.  For the
moment, we remark that Eqs.~(\ref{c}) and (\ref{a}) imply:
\begin{equation}
 a \, \zeta =\beta, \qquad
  \frac{c}{y}=  \frac{3\,\beta +\gamma}{2(\beta +\gamma)} ,
\label{cy}
\end{equation}
and that similar relations can be obviously found between any pair of
$a$, $c$, $y$ and $\zeta$. Moreover, $y'=y$, $c'=c$ in the mean field
limit $d \to \infty$.

These findings are naturally interpreted in the context of sol-gel
transition where clusters consist of bonded multifunctional monomers
and gelation corresponds to the formation of a percolating network of
crosslinked polymers~\cite{flo,deg,stauffer,adconst}.  One can
distinguish two cases: $i)$ If clusters keep their identity for all
time and never break, like in chemical (or strong) gels with permanent
bonds then Eq.~(\ref{phi_scaling}) spans the entire range $t \gtrsim
\taue $. $ii)$ When the bonds lifetime is of the order of the
relaxation time, as for physical gels, Eq.~(\ref{phi_scaling}) is
restricted to $t \lesssim \taue $, and is eventually followed by a fast
exponential decay. This latter case is the one which is more relevant
to the present context.  We focus now our attention on \MCT\ and show
that all these scaling predictions, in their mean-field limit, are met
by the ${\mathsf F}_{12}$ schematic model.

\begin{figure} 
\includegraphics[width=7.5cm]{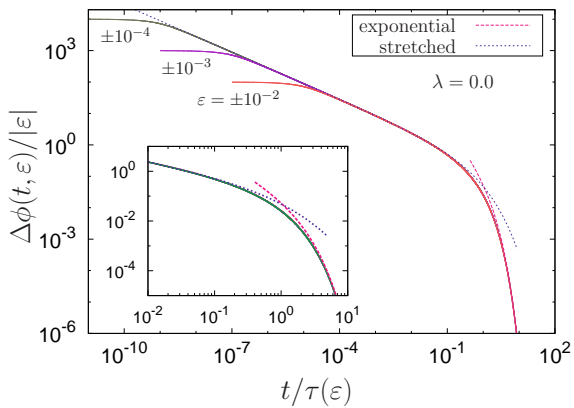} 
\caption{Non-arrested part of correlator, $\Delta \phi(t,\epsilon)$,
  vs rescaled time, $t/\taue$ for $\lambda=0$.  Full lines are the
  numerical solution of \MCT\ schematic model,
  Eqs.~(\ref{F12})-(\ref{MCT}), above ($\e>0$) and below ($\e<0$) the
  transition line. The dotted line is the stretched relaxation regime
  Eq.~(\ref{phi_scaling}) obtained by expanding the \MCT\ exact
  solution to the leading order in $t/\taue$. The dashed line
  represents the late stage exponential decay.}
\label{fig.phi_lambda0}
\end{figure}

\paragraph{Percolation approach vs  \MCT.---}

In the mean-field percolation theory one has $\beta=\gamma=1$ and the
dynamical critical exponents become:
\begin{equation}
 c = \frac{1}{x+1} , \,\,\,\, y= \frac{1}{x+1} , \,\,\,\, \zeta =2
 x,\,\,\,\, a=\frac{1}{2x},
\label{mf}
\end{equation}
from which mean-field universal relations can be derived
\begin{equation}
 \zeta \,a= 1,  \,\,\,\,\, \frac{c}{y}=1 ,  \,\,\,\,\,  c=\frac{2a}{2a+1}.   
  \label{mfrel}
\end{equation}
Notice that the first scaling relation in Eq.~(\ref{mfrel}) reproduces
the MCT scaling relation Eq.~(\ref{eq.gamma}).  Furthermore, this
percolation approach has precise predictions for the \MCT\ solutions,
namely for each value of $\lambda$, the correlator $\phi(t)$ is
described, close to the critical point, by an intermediate scaling
regime given by Eq.~(\ref{phi_scaling}), with exponents given by
Eqs.~(\ref{mf}) and $x$ determined by
\begin{equation}
 \frac{\Gamma^2(1-1/2x)}{\Gamma(1-1/x)} = \lambda .
  \label{Gamma}
\end{equation}
This last relation follows from the third relation of
Eq.~(\ref{eq.gamma}) and the fourth of Eq.~(\ref{mf}).

\begin{figure}[htbp]
\includegraphics[width=7.5cm]{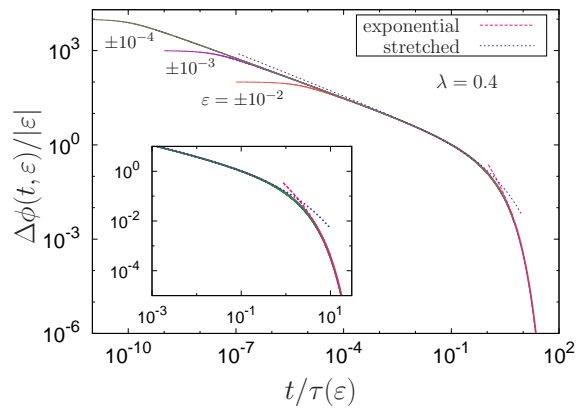}
a) 
\includegraphics[width=7.5cm]{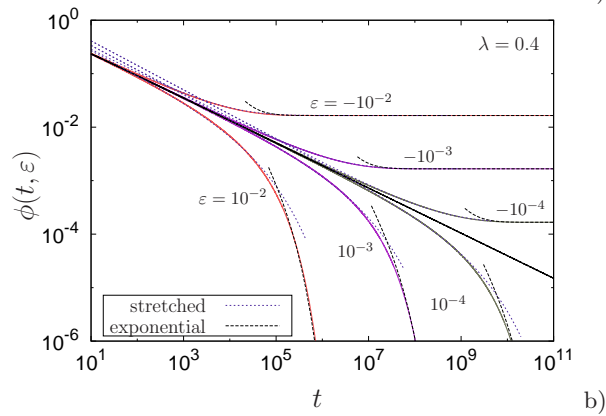} 
b)
\caption{Non-arrested part of correlator, $\Delta \phi(t,\epsilon)$ in
  scaling form (top) and in natural units (bottom) for $\lambda=0.4$.
  Full lines are the exact solution of \MCT\ schematic model,
  Eqs.~(\ref{F12})-(\ref{MCT}). The dotted line is the stretched
  relaxation regime Eq.~(\ref{phi_scaling}) with exponents determined
  according to percolation predictions Eqs.~(\ref{mf})-(\ref{Gamma})
  as $x=1.186,\, c=y=0.457,\, z=2.372$ and $a=0.422$. The dashed line
  represents the late stage exponential decay.}
\label{fig.phi_lambda04}
\end{figure}

To check the above percolation predictions against the ${\mathsf
  F}_{12}$ schematic model we first consider the simplest case
$\lambda=v_2=0$ which gives $x=1$ and, from Eq.~(\ref{mf}),
$y=c=a=1/2$ and $\zeta=2$.  In this special case, corresponding to
$v_2=0$ and $\beta=1$, the \MCT\ relaxation dynamics is exactly
known~\cite{Gotze}:
\begin{equation} 
\Delta \phi(t,\e) = \frac{|\e|}{2} \left[ \sqrt{\frac{\taue}{\pi t}}
  \exp \left(-\frac{t}{\taue} \right) - {\rm erfc}\left(
  \sqrt{\frac{t}{\taue}} \right) \right],
\label{sol.F1}
\end{equation} 
where $\Delta \phi(t,\e)$ is the non-arrested part of the correlator,
and $\taue \simeq \epsilon^{-2}$ near the transition (with
$\epsilon=1-1/v_1$). Accordingly, at short times $t/\taue \ll 1$
relaxation is algebraic, $\phi(t,\epsilon) \sim t^{-1/2}$, while at
large times, $t/\taue \gg 1$, it is exponentially fast. Expanding
Eq.~(\ref{sol.F1}) for small $t/\taue$ one finds:
\begin{eqnarray} 
\Delta \phi(t,\e) & \simeq & \frac{|\e|}{2} \sqrt{ \frac{\taue}{\pi
    t} } \left[ 1 - \sqrt{ \frac{\pi t}{\taue} } + \frac{t}{\taue}
  \right],
\end{eqnarray} 
that is, to the first leading order in $\sqrt{t/\taue}$,
Eq.~(\ref{phi_scaling}) with a normalized relaxation time
$\tilde{\taue} = \pi \taue$.  Thus the early and late stage relaxation
behaviors are bridged by a scaling regime described by
Eq.~(\ref{phi_scaling}) with exponents exactly matched by the
percolation predictions.  Fig.~\ref{fig.phi_lambda0} shows how these
three relaxation regimes compare with the exact solution of ${\mathsf
  F}_{12}$ schematic model (for $\lambda=v_2=0$).

For $\lambda>0$ the correlator cannot be expressed in a closed form.
Therefore, we have numerically solved \MCT\, Eq.~(\ref{MCT}) in the
continuous glass transition range and found an excellent agreement in
an extended region of $\lambda$ values.  In
Fig.~\ref{fig.phi_lambda04}a) we show the correlator scaling for
$\lambda=0.4$ and, to better appreciate the quality of comparison, we
replot in Fig.~\ref{fig.phi_lambda04}b) the same set of data in
natural units.  The characteristic relaxation time $\taue$ is defined
here as:
\begin{eqnarray}
\taue = \left. \int_0^{\infty} t \, \phi(t,\epsilon) \ dt \right/
\int_0^{\infty} \phi(t,\epsilon) \, dt ,
\label{tau_def}
\end{eqnarray}
and consistently reproduces the expected \MCT\ scaling $\taue \sim
|\e|^{-\zeta}$.  

The natural limits of the present description are reached when
$\lambda = v_2 \to 1$. In this case the quadratic term of the
\MCT\ kernel becomes increasingly important as compared to the linear
one and, correspondingly, the intermediate scaling regime shrinks.
This is simply understood by considering that the point $\lambda=1$
marks a crossover to a completely distinct critical behavior which is
characterized by a discontinuous ergodicity breaking.  

\begin{figure} 
\includegraphics[width=7.5cm]{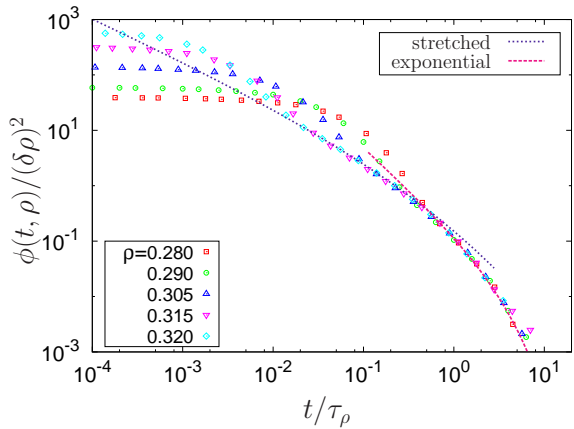} 
\caption{Rescaled persistence function $ \phi(t,\rho)$ for a
  facilitated spin model on a Bethe lattice with order parameter
  critical exponent $\beta=2$ and critical density $\rhoc=1/3$.
  Relaxation time $\tau_{\rho}$ is computed according to
  Eq.(~\ref{tau_def}) and $\delta \rho = 1 - \rho/\rhoc$.  The dotted
  line is the stretched relaxation regime Eq.~(\ref{phi_scaling}) with
  exponents determined from the measure of critical relaxation
  exponent, $a$, and Eqs.~(\ref{c})-(\ref{a}). The dashed line
  represents the late stage exponential decay.}
\label{fig.phi_scaling_f3c4}
\end{figure}

\paragraph{Percolation approach vs  cooperative facilitation.---}

To substantiate more generally the predictions of the percolation
approach to glassy systems with $\beta \neq 1$, we finally investigate
an istance of cooperative facilitated dynamics on a Bethe
lattice~\cite{SeDeCaAr}.  A set of $N$ non-interacting binary spins in
a magnetic field favoring up states evolves with a Metropolis-like
dynamics in which a randomly chosen spin is flipped if and only if at
least $f$ of its $z$ neighbors are down. For $f=3$ and $z=4$ this
facilitated dynamics undergoes a continuous ergodicity breaking at a
critical density of up spins $\rhoc=1/3$. Since the incipient cluster
of frozen spins has a fractal structure with no dangling ends the
order parameter critical exponent is $\beta=2$. We have simulated the
dynamical behavior of this model with a continuous time algorithm and
studied the persistence function $\phi(t)$ (the probability that a
spin has never flipped between times $0$ and $t$) and the relaxation
time $\tau_{\rho}$. To compare persistence data with percolation
predictions we first measure $a \simeq 0.82$ from the critical decay
at $\rhoc$. The remaining critical exponents are then inferred from
Eqs.~(\ref{c}) and (\ref{a}). Note that for this backbone percolation
problem the relevant value of the mean-field critical exponent
$\gamma$ is $\gamma= d \nu - 2 \beta = -1$.  By doing so we get
$x=\zeta =2/a \simeq 2.4, \, y \simeq 0.29$ and $c \simeq 0.73$.
Fig.~\ref{fig.phi_scaling_f3c4} shows the rescaled persistence vs
$t/\tau_{\rho}$ for several values of the density of up spins $\rho$
near the threshold.  We clearly see that also for this facilitated
dynamics, the scaling relations Eqs.~(\ref{phi1}), (\ref{phi3}) and
(\ref{phi_scaling}) provide an excellent description of relaxation
behavior.

\paragraph{Conclusions.---}

To summarise, we have established a close analogy between a dynamical
percolation approach and glassy systems with continuous glass
transition which is built upon \MCT\ as a mean-field starting point.
Consequently, \MCT\ for type A transition provides a useful mean field
approach to gelling systems, in the same way as \MCT\ for type B
transition provides a mean field framework for structural glasses.
 
Our percolation approach yields detailed predictions for the critical
exponents in any spatial dimensions and a new intermediate scaling
regime of correlation function.  Any finite dimensional generalization
of \MCT\ for systems with continuous transitions should be compared
with the above scaling laws, involving universal critical exponents of
equilibrium percolation along with a single parameter, governing the
local relaxation dynamics of finite clusters. Our framework directly
implies an upper critical dimension of 6 and a percolation critical
length for such systems.

Some of these quantitative predictions have been previously confirmed
by quasi-scattering experiments~\cite{martin} and large scale
numerical simulations~\cite{FiAbCo,cubetti} of permanent gels, and we
expect they have a much wider relevance for glassy relaxation with
continuous ergodicity breaking, including quenched-annealed
mixtures~\cite{Krak,KurzKim,Franosch}, colloidal gelation~\cite{Zac}
and vulcanization~\cite{Gold}.  It would be also interesting to
revisit in this perspective systems for which a proper definition of
cluster is rather tricky or lacking, such as random-field Ising models
and spin-glasses~\cite{Ogielski}. In those cases, $x$ could be
inferred indirectly through a measurement of global critical
relaxation and thus dynamic scaling laws should be readily tested.

Finally, it would be highly valuable to generalize the present
approach to glassy systems with a two-step relaxation scenario and
discontinuous ergodicity breaking.  Cooperative facilitation dynamics
governed by a bootstrap percolation process suggests that this may be
possible~\cite{bootstrap}.

\bibliographystyle{apsrev}

\begin{thebibliography}{50}
\bibitem{notax} The value of $x$ must be limited to
  $x>\beta/(\beta+\gamma)$ otherwise the relations with the exponents
  do not hold.
\end{thebibliography}

\end{document}